\documentclass[prl,twocolumn,english,showpacs]{revtex4-1}
\usepackage{graphicx}
\usepackage{amssymb}
\usepackage{amsmath}
\usepackage{bm}

\begin{document}

\title{Clock shift in a strongly interacting two-dimensional 
       Fermi gas}

\author{Christian Langmack$^1$, Marcus Barth$^2$, 
        Wilhelm Zwerger$^2$, and Eric Braaten$^1$}

\affiliation{
$^1${Department of Physics, The Ohio State University, 
Columbus, OH 43210, USA}
\\ 
$^2${Technische\,Universit{\"a}t\,M{\"u}nchen,\,Physik\,Department,\,
James-Franck-Strasse,\,85748 Garching,\,Germany}}

\begin{abstract}
 We derive universal relations for the radio-frequency (rf) 
spectroscopy of a two-dimensional Fermi gas consisting of two spin states 
interacting through an S-wave scattering length. 
The rf transition rate has a high-frequency tail that is 
proportional to the contact and displays logarithmic scaling violations,
decreasing asymptotically like 
$1/(\omega^2 \ln^2 \omega)$.  Its coefficient is proportional to 
$\ln^2(a_{2D}'/a_{2D})$, where $a_{2D}$ and $a_{2D}'$ are the 2-dimensional 
scattering lengths associated with initial-state and final-state interactions.
The clock shift is proportional to the contact and to $\ln(a_{2D}'/a_{2D})$.  
If $|\ln(a_{2D}'/a_{2D})| \gg 1$, the clock shift arises as a cancellation 
between much larger contributions proportional to $\ln^2(a_{2D}'/a_{2D})$ 
from bound-bound and bound-free rf transitions.
\end{abstract}


\date{\today}

\maketitle

Spectroscopy of the internal levels of atoms
is a central subject of atomic physics.
The associated transition frequencies
of a single atom are known with the precision of atomic clocks.
In fact, atomic clocks use a hyperfine transition in cesium 
to keep our standard of time. 
In a gas, the atoms experience energy shifts due to interactions 
with the surrounding atoms, resulting in 
transition frequencies that differ from those of free atoms. 
Such density-dependent interaction shifts
are a major source of systematic errors in atomic clocks, 
and are thus called {\it clock shifts}. 
Precision spectroscopy aims to avoid these shifts as far as possible, 
e.g.\ by working at extremely low densities 
or by using a purely spin-polarized gas of fermions in the ultracold limit
where the remaining S-wave collisions 
are forbidden by the Pauli principle~\cite{Gupta2003}.
Alternatively, the atoms may be stored in an optical lattice, 
where both the center-of-mass motion and the interactions
are completely quenched~\cite{JunYe2011}.  

In investigations of many-body physics using ultracold gases, 
the clock shift is, however, a signal of interest rather than a nuisance.
It can be especially interesting when the atoms are 
strongly interacting, i.e.\ their
scattering length is large compared to both the range of 
interactions and the average interparticle distance.
The interaction energy~\cite{Regal2003a, Gupta2003} 
and the pairing gap~\cite{Chin2004} have been measured
for strongly-interacting fermionic atoms near a Feshbach resonance. 
The generic setup for these experiments involves
a gas containing atoms in just two hyperfine states 
$\left|1\right>$ and $\left|2\right>$.
A radio-frequency (rf) pulse, tuned to the hyperfine splitting between 
$\left|2\right>$ and a third, unoccupied state $\left|3\right>$ 
will then transfer atoms from $\left|2\right>$ into $\left|3\right>$.
If, for example, states $\left|1\right>$ and $\left|2\right>$  
form a molecular bound state, the rf pulse has to be detuned 
compared to the bare hyperfine splitting  by a frequency $\omega$ associated 
with the binding energy of the molecule. 
At the two-body level,
effects like these can be analyzed in a straightforward manner, 
even in the presence of strong final-state interactions
between $\left|1\right>$ and $\left|3\right>$~\cite{Chin2005}. 
In a genuine many-body situation, however, 
analytic methods for calculating the rf transition rate $\Gamma(\omega)$ 
are not available. 
Exact relations for $\Gamma(\omega)$ that hold independent of density 
and interaction strength are therefore of considerable interest.

For strongly interacting fermions with two spin states 
in three dimensions (3D), a number of exact relations 
that connect thermodynamic variables to large-momentum 
and high-frequency tails of correlation functions have been derived 
by Tan~\cite{Tan2008, Braaten:2010if}. 
These relations all involve the contact $C$, which basically measures the 
probability for pairs of fermions in states $\left|1\right>$ 
and $\left|2\right>$ to be very close together. 
More precisely, the contact may be defined by the universal power-law tail 
\begin{equation}
n_{\sigma}(k) \longrightarrow C/k^{4} ,~~~~~\sigma=1,2 
\label{eq:mom-tail}
\end{equation}
of the momentum distribution of either spin state at large momentum $k$.
The Tan relations are universal in the sense that
they apply to {\it any} state of the system, 
e.g.\ few-body or many-body, homogeneous or in a trapping potential,
Fermi-liquid or superfluid state, 
provided only that the length scales associated with the temperature
and number densities are large compared to the range of interactions. 
The origin of this universality is that the Tan relations are a consequence 
of operator identities, some of which follow from
the operator product expansion of quantum field theory~\cite{Braaten:2008uh}.
There are also universal relations governing rf spectroscopy. 
Specifically, 
the average clock shift, 
i.e.\ the first moment of $\Gamma(\omega)$, is proportional 
to the contact $C$~\cite{Punk2007} or, equivalently, 
to a derivative of the total energy with respect to the scattering length~\cite{Baym2007}. 
Moreover, the rf transition rate $\Gamma(\omega)$ has a high-frequency tail 
that is proportional to $C$ and decreases like
$1/\omega^n$, where the exponent $n$ is either $\frac52$ or $\frac32$ 
depending on the strength of the final-state interactions between 
$\left|1\right>$ and 
$\left|3\right>$~\cite{Haussmann2009, Schneider2010,Braaten2010}.
Some of the universal relations have been verified experimentally,
e.g.\ by comparing the values of $C$ obtained from the tail of the 
momentum distribution in Eq.~(\ref{eq:mom-tail}) and from the 
rf transition rate at large $\omega$~\cite{Stewart2010}. 

In our present work, we derive universal relations 
for rf spectroscopy in {\it two}-dimensional  Fermi gases. 
These systems have been investigated in
several recent experiments~\cite{Kohl:1012, Kohl:1110, Zwierlein:1110}. 
Beyond the motivation provided by these experiments, universal 
relations that constrain the rf transition rate in two dimensions (2D) 
are of interest also for other reasons.
First, rf spectroscopy measurements can be performed within a 
rather short time scale. It thus allows access to not only equilibrium 
but also excited states of the many-body system, 
e.g.\ the polaron on the repulsive branch of a Feshbach 
resonance~\cite{Schmidt2011} or Bose gases in the regime of strong repulsion.
A second reason is that a 2D gas with zero-range interactions provides 
an example of a non-relativistic many-body system that is scale invariant 
at the classical but not at the quantum level. 
This breaking of the scale invariance by quantum effects is
an elementary example of an anomaly~\cite{Holstein1993}. 
It implies that the coupling constant $g$ 
(which will be defined explicitly below) is in fact 
a running coupling constant $g(\mu)= - 2\pi/\ln (a_{2D} \mu)$ 
that changes logarithmically with the momentum scale $\mu$,
reminiscent of the coupling constant of Quantum Chromodynamics.
The associated intrinsic length $a_{2D}$ sets the scale for
the two-body scattering amplitude which, at low energies, has the 
characteristic form~\cite{Petrov2001, Bloch2008}
\begin{equation}
 f(q)= \frac{4\pi}{\ln(1/q^2 a_{2D}^2)+i\pi} \,.
\label{coupling}
\end{equation}
For any positive value of $a_{2D}$, this amplitude has a pole at $q=i/a_{2D}$,
which signifies the presence of a diatomic molecule with binding energy 
$E_d = \hbar^2/m a_{2D}^2$ that we will call the (12)-dimer.

In experiments with ultracold atoms, 
the 2D gas arises from a 3D system by adding a strong 
transverse confining potential that restricts the atoms to the ground state  
in the third dimension.  For harmonic trapping with angular frequency 
$\omega_z$, the relation between $a_{2D}$ and the experimentally tunable 
3D scattering length $a$
is given by $\ell/a=f_1(\ell^2/a_{2D}^2)$, 
where $\ell^2 = \hbar/m \omega_z$ and the function $f_1(x)$ is 
given in section V.A of Ref.~\cite{Bloch2008}.

A system consisting of fermions with the two spin states 
$\left|1\right>$ and $\left|2\right>$
can be described by a quantum field theory with 
the interaction Hamiltonian 
\begin{equation}
H_{\rm int}=
(g/m)\mbox{$\int$} d^2R \; \psi^\dagger_1\psi^\dagger_2 \psi_2 \psi_1(\bm{R}).
\label{equ:hamiltonian}
\end{equation}
Here $g$ is the bare coupling constant, which can be identified with the
running coupling constant at the momentum scale of the ultraviolet cutoff.
Within this framework, universal relations for both equilibrium properties 
and rf spectra can be derived using the
operator product expansion (OPE) \cite{Braaten:2008uh}.
In particular, the tail of the momentum distribution 
in Eq.~(\ref{eq:mom-tail}) follows from the OPE for 
$\psi_{\sigma}^{\dagger} (\bm{R}+\frac12 \bm{r}) 
\psi_{\sigma} (\bm{R}-\frac12 \bm{r})$. 
The leading contribution that is non-analytic as $\bm{r} \to 0$
comes from the operator 
$\psi_1^{\dagger} \psi_2^{\dagger} \psi_2 \psi_1 (\bm{R})$,
whose coefficient exhibits a logarithmic singularity in 2D
of the form $|\bm{r}|^2 \log|\bm{r}|$. As a result, 
$n_\sigma(k)$ behaves asymptotically like $C/k^4$ 
with no logarithmic corrections, as in 3D and also in 1D~\cite{Barth2011}.
The contact $C=\int d^2R\,\,{\cal C}(\bm{R})$ is an integral of
the contact density
\begin{equation}
{\cal C}(\bm{R})  = 
g^2 \langle \psi_{1}^{\dagger} \psi_{2}^{\dagger} 
			\psi_2 \psi_1 (\bm{R})\rangle\, ,
\label{eq:contact-opdef}
\end{equation} 
which is the expectation value of a local operator
that annihilates and recreates a pair of fermions. The structure of 
these results is identical to the 3D case and, indeed, 
many of Tan's universal relations 
may be generalized to 2D~\cite{werner2010}.
Universal relations for current correlators, structure factors, 
and hydrodynamic correlators in 2D
have also been derived using the OPE~\cite{hofmann2011}.

We now consider the rf spectroscopy of the 2D system
consisting of fermions in states $\left|1\right>$ and $\left|2\right>$. 
Beyond the interaction 
described by $a_{2D}$, we also include final-state interactions 
between $\left|1\right>$ and $\left|3\right>$ with 2D scattering length $a_{2D}'$.
We will refer to the diatomic molecule consisting of $\left|1\right>$ 
and $\left|3\right>$, which has
binding energy $E_d' = \hbar^2/m a_{2D}'{}^2$, as the (13)-dimer.
The transfer rate $\Gamma(\omega)$ 
between $\left|2\right>$ and $\left|3\right>$ can be expressed 
in terms of the imaginary part of a time-ordered correlation function:
\begin{equation}
\label{eq:rftransgen}
\Gamma(\omega)  =	\Omega^2 \; \mbox{$\int$} d^2 R ~\textrm{Im} \left(i 
			\left\langle
                	{\cal O}_{\rm rf}(\omega+i \epsilon,\bm{R}) 
			\right\rangle\right),
\end{equation}
where $\Omega$ is the Rabi frequency of the rf transition and
${\cal O}_{\rm rf}(\omega,\bm{R})$ is an operator that depends on a 
complex variable $\omega$:
\begin{eqnarray}
{\cal O}_{\rm rf}(\omega,\bm{R}) &=&
			\mbox{$\int$} dt \;  e^{i \omega t} \mbox{$\int$} d^2 r \; \nonumber \\ 
\label{eq:rfamp}
& & \hspace{-1.5cm}\times {\cal T}
                	\psi_{2}^{\dagger} \psi_{3}
			\left(\bm{R}+\mbox{$\frac{1}{2}$}\bm{r},t\right)
                	\psi_{3}^{\dagger} \psi_{2} 
			\left(\bm{R}-\mbox{$\frac{1}{2}$}\bm{r},0\right).
\end{eqnarray}
(The time-ordering symbol ${\cal T}$ acts on all operators to the 
right.)  The bilocal operator can be expanded in terms of 
local operators by using the OPE:
\begin{equation}
\label{eq:ope}
{\cal O}_{\rm rf}(\omega,\bm{R}) 
	= \mbox{$\sum$}_{n}{C_n(\omega) {\cal O}_{n}(\bm{R})}.
\end{equation}
Local operators can be assigned scaling dimensions. The quantum fields
$\psi_{\sigma}$ have dimension 1. The gradient $\mathbf{\nabla}$ 
and the time derivative $\partial/ \partial t$ increase the 
dimension by 1 and 2, respectively. 
At large frequency $\omega$, higher dimension operators in the OPE 
have Wilson coefficients 
$C_{n}(\omega)$ that decrease with higher powers of $\omega$.
The Wilson coefficients $C_{n}(\omega)$ in Eq.~\eqref{eq:ope} can be calculated 
by matching Green functions of the bilocal rf operator on the left side 
of the OPE with Green functions of the local operators on the right
side.  For operators of dimensions 2 and 4, it is  
sufficient to match the Green functions in the 1-atom and 2-atom sectors.  
Our final result for the expectation 
value of ${\cal O}_{\rm rf}(\omega,\bm{R})$ at large complex $\omega$, 
including all operators with dimensions up to 4, is
\begin{eqnarray}
& \langle {\cal O}_{\rm rf}(\omega,\bm{R}) \rangle &
 	\longrightarrow 	
	\frac{i}{\omega} n_{2} (\bm{R})
  + \frac{i \; \nabla^{2} n_2 (\bm{R}) + 4 m \mathbf{\nabla} \cdot \bm{J}_{2}(\bm{R})}{8 m \omega^2}
\nonumber \\
& & + \frac{i \ln(a_{2D}'/a_{2D}) \ln(a_{2D} \sqrt{-m \omega}\,)}
         {2 \pi m \omega^2 \ln(a_{2D}' \sqrt{-m \omega}\,)} 
	{\cal C}(\bm{R}),
\label{eq:rfope}
\end{eqnarray}
where $n_2$ and $\bm{J}_2$ are the number density and current density 
for $\left|2\right>$.  If the system is in a steady state, 
$\mathbf{\nabla} \cdot \bm{J}_2$ vanishes because of current conservation.
The integral of $\nabla^2 n_2$ vanishes by the divergence theorem.
The terms shown explicitly in Eq.~(\ref{eq:rfope})
are not affected by final-state interactions between 
$\left|2\right>$ and $\left|3\right>$, which first enter into the 
coefficients of dimension-6 operators.

We can derive various universal relations from the OPE in Eq.~\eqref{eq:rfope}. 
The tail of the rf transition rate at large positive frequencies $\omega$ 
can be obtained by extrapolating Eq.~\eqref{eq:rfope} to the real axis,
taking the imaginary part, and then inserting 
it into Eq.~\eqref{eq:rftransgen}:
\begin{equation}
\Gamma(\omega)  \longrightarrow 
		     \frac{\ln^2(E_d'/E_d) \Omega^2 C}
		     {4 m \omega^2 \big[
		     \ln^2(\omega/E_d') + \pi^2 \big]}  .
\label{eq:rftail}
\end{equation}
Higher dimension operators in the OPE give contributions to the
high-frequency tail that decrease faster than $\omega^{-3}$.
These contributions are suppressed by a positive power of
an energy scale of the many-body system divided by $\omega$.
The relevant energy scales of the many-body system 
include $E_d$, the thermal energy $k_B T$, 
and the two Fermi energies $4 \pi^2 n_1/m$ and $4 \pi^2 n_2/m$.
The factor $\ln^2(E_d'/E_d)$ in Eq.~(\ref{eq:rftail})
can be interpreted as $16 \pi^2/g^2(1/a_{2D}')$, 
where $g(1/a_{2D}')$ is the running coupling constant 
at the momentum scale $1/a_{2D}'$. 
If $a_{2D}'=a_{2D}$,
the high-frequency tail in Eq.~\eqref{eq:rftail} vanishes identically. 
In this case, $\Gamma(\omega)$ just consists of an unshifted peak
$\pi \Omega^2 N_{2} \delta(\omega)$, because the rf pulse merely 
rotates $\left|2\right>$ and $\left|3\right>$ in spin space~\cite{Baym2007}.
The nontrivial dependence on $\omega$ in Eq.~\eqref{eq:rftail}
is a consequence of the anomalous scale invariance. 
If $|\ln(\omega/E_d')| \ll \pi$, the high-frequency tail 
in Eq.~(\ref{eq:rftail}) has the simple scaling behavior  
$\Gamma(\omega) \sim 1/\omega^2$. However at much larger or much smaller
frequencies, there are logarithmic scaling violations and the asymptotic
behavior is $\Gamma(\omega)\sim (\omega^2\ln^2\omega)^{-1}$.
Simple scaling behavior also arises in the limit of
negligible final-state interactions in which 
$\ln(E_d'/\omega) \rightarrow \infty$.
In this limit, the high-frequency tail in Eq.~(\ref{eq:rftail}) reduces to 
$\Omega^2 C / 4 m \omega^2$. 
This relation has been used in recent experiments with $^{40}\text{K}$ atoms
to extract the contact density in the normal state of the 2D Fermi gas~\cite{Frohlich2011}. 
The contact density and its dependence on 
$a_{2D}$ is close to that obtained
from variational Monte Carlo calculations at $T=0$~\cite{Bertaina2011}.

Sum rules can be derived by expressing weighted integrals of $\Gamma(\omega)$ 
as contour integrals in the complex-$\omega$ plane:
\begin{equation}
\int_{-\infty}^{+\infty} \hspace{-0.6 cm} d \omega  f(\omega) 
\Gamma(\omega) = \frac{\Omega^2}{2} 
\int  \!\! d^2 R \oint d \omega f(\omega) \langle
                	{\cal O}_{\rm rf}(\omega,\bm{R}) 
\rangle,
\label{eq:contourint}
\end{equation}
where the $\omega$  contour runs from $+\infty$ to $-\infty$ just below the real 
axis and then from $-\infty$ to $+\infty$ just above the axis.  
If the contour is deformed into a circle of infinite radius, the OPE in 
Eq.~\eqref{eq:rfope} will be exact everywhere on the contour.  
Choosing $f(\omega)=1$ and 
$f(\omega)=\omega$, we obtain
\begin{subequations}
\begin{eqnarray}
\int_{-\infty}^{+\infty} \hspace{-0.5 cm} d \omega \; \Gamma(\omega) 
& = &\pi \Omega^{2} N_{2}, 
\label{eq:sumrule0}
\\
\int_{-\infty}^{+\infty} \hspace{-0.5 cm} d \omega \; \omega \Gamma(\omega) 
& = & \frac{\ln(a_{2D}'/a_{2D}) \Omega^2 C}{2 m} .
\label{eq:sumrule1}
\end{eqnarray}
\label{eq:sumrules}
\end{subequations}

\vspace{-12pt}\noindent
These sum rules are exact: they receive no contributions 
from higher dimension operators in the OPE.
The first sum rule guarantees that interactions do 
not change the integrated rf signal.
The second sum rule determines the clock shift $\langle \omega \rangle$,
which is defined as
the ratio of the integrals in Eqs.~\eqref{eq:sumrules}:
\begin{equation}
\langle \omega \rangle = 
- \frac{\ln(E_d'/E_d)\, C}{4 \pi m N_2}.
\label{eq:clock}
\end{equation}
Note that both the clock shift and the tail of the rf transition rate 
in Eq.~\eqref{eq:rftail} are unaffected by final-state interactions
 between $\left|2\right>$ and $\left|3\right>$.

\begin{figure}[t]
\includegraphics*[width=3.2in,angle=0]{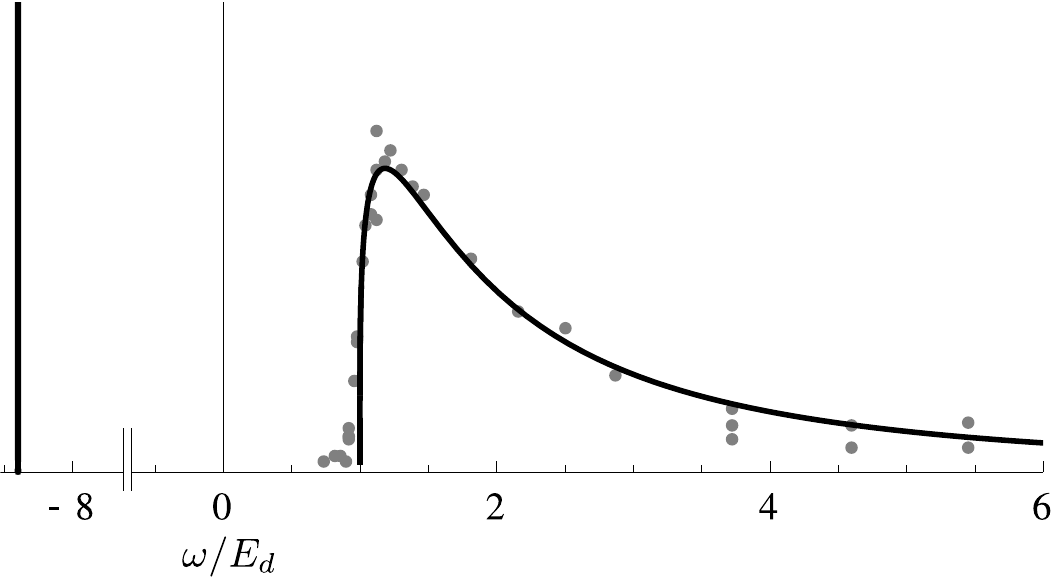}
\caption{
Rf transition rate for the (12)-dimer as a function of 
$\omega/E_d$ for the case 
$E_d' = 9.4 E_d$. 
There is a delta function at $\omega = - (E_d'-E_d)$
from the bound-bound transition.
The threshold for bound-free transitions is at $\omega = E_d$. 
The grey dots in the bound-free region are data from 
Zwierlein and collaborators \cite{Zwierlein:pc}, 
normalized by a least-squares fit to the theory curve. }
\label{fig:rfdimer}
\end{figure}

The universal relations in Eqs.~\eqref{eq:rftail} and \eqref{eq:sumrules}
can be illustrated by considering rf spectroscopy on the two-body system
consisting of a single (12)-dimer with binding energy $E_d$.
The rf transition rate is
\begin{eqnarray}
\Gamma(\omega) & = & 
\frac{\pi \Omega^2 E_d'E_d \ln^2(E_d'/E_d)}{(E_d'-E_d)^2}~
\delta(\omega - E_d + E_d') 
\nonumber \\
&& \hspace{-0.35cm} 
+ \frac{\pi \Omega^2 E_d \ln^2(E_d'/E_d)}
        {\omega^2 [\ln^2((\omega-E_d)/E_d') + \pi^2]}~\theta(\omega - E_d).
\label{eq:rf-dimer}
\end{eqnarray}
By evaluating the integrals in Eqs.~\eqref{eq:sumrules},
we verify that the sum rules are satisfied exactly with $N_2=1$ 
and $C = 4 \pi m E_d$, the contact for the (12)-dimer
obtained from the adiabatic universal relation 
$dE_d/da_{2D} = - C/(2 \pi m a_{2D})$ \cite{werner2010}.
The universal relation for the tail in Eq.~\eqref{eq:rftail}
holds for $|\omega| \gg E_d$. 
The clock shift for the (12)-dimer has a negative contribution 
proportional to $\ln^2(E_d'/E_d)$ from bound-bound transitions.
If $|\ln(E_d'/E_d)| \gg 1$, there is a large canceling contribution
from bound-free transitions, such that  
the total clock shift agrees with Eq.~\eqref{eq:clock}.
Near the threshold, the bound-free transition rate has the behavior
$1/\ln^2((\omega-E_d)/E_d')$.
This follows from the inverse-logarithmic behavior of the 
two-body scattering amplitude for $\left|1\right>$ and $\left|3\right>$
analogous to Eq.~\eqref{coupling}.
Recent experiments using $^6\text{Li}$ atoms in 2D~\cite{Zwierlein:1110}
are in good agreement with the bound-free transition rate 
in Eq.~\eqref{eq:rf-dimer}
for a single (12)-dimer in the presence of final-state interactions.
The data in Fig.~\ref{fig:rfdimer}~\cite{Zwierlein:pc} were measured near the 
3D Feshbach resonance for $\left|1\right>$ and $\left|2\right>$
at a magnetic field of 690.7~G 
and a lattice depth to recoil energy ratio of 18.6(7),
and they extend out to higher $\omega$ than in Ref.~\cite{Zwierlein:1110}.
The ratio $E_d'/E_d$~ is around $9.4$
and $E_d$ is about a factor $6$ larger than the Fermi energy~\cite{Zwierlein:pc}.
This places the system in the BEC limit of the many-body
problem where the two-body result in Eq.~\eqref{eq:rf-dimer} applies.  

We now consider the many-body system in the limit in which $E_d'$ 
is much greater than $E_d$ and the other relevant energy scales. 
The bound-bound and free-bound region is then well separated from 
the free-free and free-bound region of the rf spectrum. 
There could be additional structure in the region of $\omega$
between $-(E_d'-E_d)$ and 0 associated 
with clusters of three or more atoms. 
However, if this region is featureless, it is useful to define a clock shift
for the rf transition rate restricted to a range of frequencies
$-\omega_0 \hspace{-3pt} < \hspace{-2pt} \omega \hspace{-2pt} < \hspace{-3pt} \omega_0$ that excludes the bound-bound peak.
We choose $\omega_0$ and $E_d'-\omega_0$ to be much greater than 
the energy scales of the system.  
This clock shift is given by a ratio of integrals analogous to those in
Eq.~(\ref{eq:sumrules}).  They can be expressed as contour integrals
analogous to those in Eq.(\ref{eq:contourint}), except that the 
contour runs from $+\omega_0$ to $-\omega_0$ 
just below the real axis and then back to $+\omega_0$ 
just above the axis. If the contour is deformed 
into a circle of radius $\omega_0$, we can apply the OPE in 
Eq.~\eqref{eq:rfope} and obtain the sum rules
\begin{subequations}
\begin{eqnarray}
\int_{-\omega_0}^{+\omega_0} 
\hspace{-0.5 cm} d \omega \; \Gamma(\omega) 
& = & \pi \Omega^{2} N_{2} 
- [{\rm Im} \, {\rm Ei}(-L_0 + i \pi)] 
\nonumber \\
 && 
\times \left( \ln \frac{a_{2D}'}{a_{2D}} \right)
\frac{\Omega^2 C}{\pi m E_d'} + \ldots, 
\label{eq:FEsumrule0}
\\
\int_{-\omega_0}^{+\omega_0} 
\hspace{-0.5 cm} d \omega \; \omega \Gamma(\omega) 
& = & \left( \ln \frac{a_{2D}'}{a_{2D}}  
- \frac{2}{\pi} \ln^2 \frac{a_{2D}'}{a_{2D}} \arctan \frac{\pi}{L_0} \right)
 \frac{\Omega^2 C}{2 m} \nonumber \\ & &  + \ldots ,
\label{eq:FEsumrule1}
\end{eqnarray}
\label{eq:FEsumrules}
\end{subequations}

\vspace{-12pt} \noindent where ${\rm Ei}(x)$ is the exponential integral function
and $L_0 = \ln(\omega_0/E_d')$.
One advantage of finite-frequency sum rules such as those 
in Eqs.~(\ref{eq:FEsumrules})
is that they can be evaluated using measured rf spectra
without extrapolations to larger frequencies.
As indicated by ``$+ \ldots$'' in Eqs.~(\ref{eq:FEsumrules}), 
these sum rules are not exact.
They receive contributions from higher dimensional operators
in the OPE that are suppressed by powers of 
an energy scale of the system divided by 
$\omega_0$ or $E_d'$.  For example, the higher dimension 
contribution in Eq.~(\ref{eq:FEsumrule0})
is suppressed by a factor of $C/m E_d' N_2$.
The corrections that are suppressed by inverse powers of $\omega_0$
can be minimized by choosing $\omega_0$ as large as possible 
while still avoiding contributions from the bound-bound peak.
Neglecting the contributions from higher dimension operators,
the clock shift obtained from Eqs.~(\ref{eq:FEsumrules}) reduces to
\begin{equation}
\langle \omega \rangle = 
\left( \frac{1}{\pi} \ln^2 \frac{E_d'}{E_d} \arctan \frac{\pi}{|L_0|} 
- \ln \frac{E_d'}{E_d}\right)
\frac{C}{4 \pi m N_2}\,.
\label{eq:FEclock}
\end{equation}
If $|\ln(E_d'/E_d)|\gg 1$, this clock shift is much larger than that 
in Eq.~(\ref{eq:clock}).
The clock shift in Eq.~(\ref{eq:FEclock}), which excludes the 
contribution from bound-bound transitions,
is the one that is more relevant 
to the limit of weak final-state interactions.
The factor $\ln^2(E_d'/E_d)$ diverges in the limit $a_{2D}' \to 0$,
indicating that the clock shift in this case is quadratically 
logarithmically sensitive to the range of interactions.

In summary,
we have presented universal relations for the rf spectroscopy of strongly
interacting fermions in a 2D Fermi gas.
They are of direct relevance to current experiments with ultracold atoms,
providing rigorous connections between rf spectrocopy, thermodynamics, 
and various correlation functions. 
They reveal aspects of these many-body systems 
that are directly related to the breaking of scale invariance in 2D
and will hopefully provide inspiration for further 
investigations of these unusual systems.

This research was supported in part by a joint grant from the 
ARO and the AFOSR and by the DFG.
We acknowledge valuable discussions with 
M.~K\"ohl, R.~Schmidt, A.~Sommer, and M.~Zwierlein
and we thank the latter two for making their data available.


\begin{thebibliography}{10}%

\bibitem{Gupta2003}
S.~Gupta et al.,
Science {\bf 300}, 1723 (2003).

\bibitem{JunYe2011}
M.~Swallows et al.,
Science {\bf 331}, 1043 (2011).

\bibitem{Regal2003a}%
C.~A.~Regal and D.~S.~Jin,
Phys.\ Rev.\ Lett.\ {\bf 90}, 230404 (2003).

\bibitem{Chin2004}
C.~Chin et al.,
Science {\bf 305}, 1128 (2004).

\bibitem{Chin2005}
C.~Chin and P.~S.~Julienne,
Phys.\ Rev.\ A {\bf 71}, 012713 (2005).

\bibitem{Tan2008}
S.~Tan,
Ann.\ Phys.\ (N.Y.)\ {\bf 323}, 2952 (2008);
\ {\bf 323}, 2971 (2008); 
\ {\bf 323}, 2987 (2008).

\bibitem{Braaten:2010if}
E.~Braaten,
in {\it The BCS-BEC Crossover and the Unitary Fermi Gas},  
edited by\ W.~Zwerger
(Springer-Verlag, 2012).
  
\bibitem{Braaten:2008uh}
  E.~Braaten and L.~Platter,
  Phys.\ Rev.\ Lett.\  {\bf 100}, 205301 (2008).

\bibitem{Punk2007}
M.~Punk and W.~Zwerger,
Phys.\ Rev.\ Lett.\ {\bf 99}, 170404 (2007).

\bibitem{Baym2007}
G.~Baym, C.J.~Pethick, Z.~Yu, and M.W.~Zwierlein,
Phys.\ Rev.\ Lett.\ {\bf 99}, 190407 (2007).

\bibitem{Haussmann2009}
R.~Haussmann, M.~Punk, and W.~Zwerger,
Phys.\ Rev.\ A {\bf 80}, 063612 (2009).

\bibitem{Schneider2010}
W.~Schneider and M.~Randeria,
Phys.\ Rev.\ A {\bf 81}, 021601 (2010).

\bibitem{Braaten2010}
E.~Braaten, D.~Kang, and L.~Platter,
Phys.\ Rev.\ Lett.\ {\bf 104}, 223004 (2010).

\bibitem{Stewart2010}
J.T.~Stewart, J.P.~Gaebler, T.E.~Drake, and D.S.~Jin,
Phys.\ Rev.\ Lett.\ {\bf 104}, 235301 (2010).

\bibitem{Kohl:1012}
B.~Fr{\"o}hlich et al.,
Phys.\ Rev.\ Lett.\ {\bf 106}, 105301 (2011).

\bibitem{Kohl:1110}
M.~Feld et al.,
Nature (London) {\bf 480}, 75 (2011).

\bibitem{Zwierlein:1110}
A.~Sommer et al.,
Phys\ Rev.\ Lett.\ {\bf 108}, 045302 (2012).

\bibitem{Schmidt2011}
R.~Schmidt, T.~Enss, V.~Pietil\"{a}, and E.~Demler,
Phys.\ Rev.\ A\ {\bf 85}, 021602(R) (2012).

\bibitem{Holstein1993}
B.~Holstein,
Am.\ J.\ Phys.\ {\bf 61}, 142 (1993).
  
\bibitem{Petrov2001}
D.S.~Petrov and G.V.~Shlyapnikov,
Phys.\ Rev.\ A {\bf 64} 012706 (2001).

\bibitem{Bloch2008}
I.~Bloch, J.~Dalibard, and W.~Zwerger,
Rev.\ Mod.\ Phys.\ {\bf 80}, 885 (2008).

\bibitem{Barth2011}
M.~Barth and W.~Zwerger,
Ann.\ Phys.\ {\bf 326}, 2544 (2011).

\bibitem{werner2010}
F.~Werner and Y.~Castin,
arXiv:1001.0774.

\bibitem{hofmann2011}
J.~Hofmann,
Phys.\ Rev.\ A.\ {\bf 84}, 043603 (2011).

\bibitem{Frohlich2011}
B.~Fr{\"o}hlich, Ph.D.\ thesis,
University of Cambridge, July 2011.

\bibitem{Bertaina2011}
G.~Bertaina and S.~Giorgini, 
Phys.\ Rev.\ Lett.\ {\bf 106}, 110403 (2011).

\bibitem{Zwierlein:pc}
A.~Sommer and M.~Zwierlein, private communication.


\end{thebibliography}
\end{document}